\documentclass[10pt,letterpaper]{article}
\usepackage[top=0.85in,left=2.75in,footskip=0.75in]{geometry}

\usepackage{amsmath,amssymb}

\usepackage{changepage}

\usepackage[utf8x]{inputenc}

\usepackage{textcomp,marvosym}

\usepackage{cite}

\usepackage{nameref,hyperref}

\usepackage[right]{lineno}

\usepackage{microtype}
\DisableLigatures[f]{encoding = *, family = * }

\usepackage[table]{xcolor}

\usepackage{array}

\newcolumntype{+}{!{\vrule width 2pt}}

\newlength\savedwidth

\raggedright
\usepackage[aboveskip=1pt,labelfont=bf,labelsep=period,justification=raggedright,singlelinecheck=off]{caption}

\makeatletter
\renewcommand{\@biblabel}[1]{\quad#1.}
\makeatother

\usepackage{lastpage,fancyhdr,graphicx}
\usepackage{epstopdf}
\fancyheadoffset[L]{2.25in}
\fancyfootoffset[L]{2.25in}
\usepackage{xfrac}
\usepackage{graphicx, color}
\usepackage{booktabs}

\DeclareMathOperator{\logit}{logit}

\begin{document}
\vspace*{0.2in}

\title{}
\begin{flushleft}
{\Large
\textbf\newline{Estimating and interpreting secondary attack risk:\\ Binomial considered harmful} % Please use "sentence case" for title and headings (capitalize only the first word in a title (or heading), the first word in a subtitle (or subheading), and any proper nouns).
}
\newline
\\
Yushuf Sharker\textsuperscript{1},
Eben Kenah\textsuperscript{2, *}
\\
\bigskip
\textbf{1} Division of Biometrics, Center for Drug Evaluation and Research, Food and Drug Administration, Silver Spring, Maryland, USA
\\
\textbf{2} Biostatistics Division, College of Public Health, The Ohio State University, Columbus, Ohio, USA
\\
\bigskip

% Use the asterisk to denote corresponding authorship and provide email address in note below.
* kenah.1@osu.edu

\end{flushleft}

\section*{Abstract}
% less than 300 words
The household secondary attack risk (SAR), often called the secondary attack rate or secondary infection risk, is the probability of infectious contact from an infectious household member $A$ to a given household member $B$, where we define infectious contact to be a contact sufficient to infect $B$ if he or she is susceptible.
Estimation of the SAR is an important part of understanding and controlling the transmission of infectious diseases.
In practice, it is most often estimated using binomial models such as logistic regression, which implicitly attribute all secondary infections in a household to the primary case.
In the simplest case, the number of secondary infections in a household with $m$ susceptibles and a single primary case is modeled as a binomial($m, p$) random variable where $p$ is the SAR.
Although it has long been understood that transmission within households is not binomial, it is thought that multiple generations of transmission can be safely neglected when $p$ is small. 
We use probability generating functions and simulations to show that this is a mistake.
The proportion of susceptible household members infected can be substantially larger than the SAR even when $p$ is small. 
As a result, binomial estimates of the SAR are biased upward and their confidence intervals have poor coverage probabilities even if adjusted for clustering. 
Accurate point and interval estimates of the SAR can be obtained using longitudinal chain binomial models or pairwise survival analysis, which account for multiple generations of transmission within households, the ongoing risk of infection from outside the household, and incomplete follow-up.
We illustrate the practical implications of these results in an analysis of household surveillance data collected by the Los Angeles County Department of Public Health during the 2009 influenza A (H1N1) pandemic.

\section*{Author summary}
% between 150 and 200 words
The household secondary attack risk (SAR), often called the secondary attack rate or secondary infection risk, is the probability of infectious contact from an infectious household member $A$ to a given household member $B$, where we define infectious contact to be a contact sufficient to infect $B$ if he or she is susceptible.
The most common statistical models used to estimate the SAR are binomial models such as logistic regression, which implicitly assume that all secondary infections in a household are infected by the primary case.
Here, we use analytical calculations and simulations to show that estimation of the SAR must account for multiple generations of transmission within households.
As an example, we show that binomial models and statistical models that account for multiple generations of within-household transmission reach different conclusions about the household SAR for 2009 influenza A (H1N1) in Los Angeles County, with the latter models fitting the data better.
In an epidemic, accurate estimation of the SAR allows rigorous evaluation of the effectiveness of public health interventions such as social distancing, prophylaxis or treatment, and vaccination.

%\linenumbers
  
\section{Introduction}
In infectious disease epidemiology, the household secondary attack risk (SAR) is the probability of infectious contact from an infected household member $A$ to a susceptible household member $B$ during $A$'s infectious period, where we define infectious contact as a contact sufficient to infect $B$ if he or she is susceptible.
It is often called the secondary attack \emph{rate}, but we prefer to call it a risk because it is a probability~\cite{morgenstern1980measures}.
SARs can also be defined in other groups of close contacts, such as schools or hospital wards~\cite{de1981meningococcal}.

The SAR is used to assess the transmissibility of disease and to evaluate control measures~\cite{fox1974family, elveback1976influenza, monto1994studies, halloran2003estimating, terry2014field}. 
The idea was originally developed by Charles V. Chapin in 1903 to study the transmission of diphtheria and scarlet fever, and it was extended to influenza, tuberculosis, and other infectious diseases by Wade Hampton Frost~\cite{frost1938familial, wilson1939measles, terris1999charles}. 
Household surveillance data from emerging infections is often used to estimate the SAR, including 1957 and 1968 pandemic influenza~\cite{jordan1958study, chin1960morbidity, davis1970hong}, meningococcal disease~\cite{de1981meningococcal}, pertussis~\cite{halloran2003estimating}, SARS coronavirus~\cite{goh2004secondary}, seasonal influenza~\cite{cauchemez2004bayesian, tsang2014association, tsang2015influenza, petrie2017application}, rotavirus~\cite{banerjee2008evidence}, 2009 pandemic influenza A (H1N1)~\cite{yang2009transmissibility, cauchemez2009household, morgan2010household, france2010household, carcione2011secondary, savage2011assessing, ng2017association}, MERS coronavirus~\cite{drosten2014transmission, arwady2016middle}, Ebola virus disease~\cite{fang2016transmission, glynn2018variability, reichler2018household}, norovirus~\cite{marsh2018unwelcome, tsang2018transmissibility}, hand-foot-and-mouth disease~\cite{hoang2019transmission}, cryptosporidium~\cite{korpe2019case}, measles~\cite{banerjee2019containing}, and COVID-19~\cite{bi2020epidemiology, li2020characteristics}.

It has been understood that within-household transmission is not binomial since the work of En'ko in 1899~\cite{dietz1988first}, Reed and Frost in 1928~\cite{becker1989analysis}, and Greenwood in 1931~\cite{greenwood1931statistical}. 
The process is binomial only if the primary case (the first infected household member~\cite{giesecke2014primary}) is the only possible source of infection for susceptible household members throughout his or her infectious period.
However, binomial models continue to be used for the estimation of the SAR because it is thought that multiple generations of transmission within households can be safely neglected when the SAR is small.
In its simplest form, this assumes that the number of secondary infections in a household with $m$ susceptible individuals and a single primary case is a binomial$(m, p)$ random variable, where $p$ is the household SAR. 
A given transmission chain of length $k$ from a primary case $A$ to a given susceptible $B$ has probability $p^k$, which decays exponentially as $k$ increases.
Up to and including the COVID-19 pandemic, the vast majority of studies of household transmission use a binomial model (often a logistic regression model) to estimate the household SAR~\cite{wilson1939measles, jordan1958study, chin1960morbidity, davis1970hong, de1981meningococcal, halloran2003estimating, banerjee2008evidence, france2010household, morgan2010household, carcione2011secondary, savage2011assessing, terry2014field, drosten2014transmission, arwady2016middle, ng2017association, glynn2018variability, reichler2018household, marsh2018unwelcome, korpe2019case, banerjee2019containing, bi2020epidemiology, li2020characteristics}. 
A smaller number of studies have used explicit statistical models of transmission~\cite{cauchemez2004bayesian, yang2009transmissibility, cauchemez2009household, tsang2014association, tsang2015influenza, lau2015inferring, fang2016transmission, petrie2017application, tsang2018transmissibility}.
Here, we hope to establish that the latter approach should become universal.

Although the probability each given transmission chain of length $k$ from $A$ to $B$ decays as $p^k$, the risk of infection through $k$ generations of transmission also depends on the number of transmission chains of length $k$.
A transmission chain of length $k \geq 1$ from $A$ to $B$ can be specified by choosing $k - 1$ individuals from the $m - 1$ susceptible household members other than $B$. 
Each ordering of these $k - 1$ individuals produces a unique transmission chain.
For $1 \leq k \leq m$, the total number of paths from $A$ to $B$ of length $k$ equals the number of permutations of $k - 1$ objects chosen from $m - 1$ objects:
\begin{equation}
  P(m - 1, k - 1) = \frac{(m - 1)!}{(m - k)!}.
\end{equation}
Table~\ref{tab:paths} shows that the number of paths of length $k$ can grow quickly with household size. 
Each path can carry infection from $A$ to $B$, so the total risk of transmission from $A$ to $B$ along any path of length $k$ can be much greater than $p^k$.
A binomial model attributes this additional risk of infection to direct transmission from the primary case, so the estimated SAR is too high.

\begin{table}
    \centering
    \begin{tabular}{rrrrrr}
        \toprule
        \multicolumn{1}{c}{Susceptibles} & \multicolumn{4}{c}{Path length ($k$)} \\
        \multicolumn{1}{c}{($m$)} & 1 & 2 & 3 & 4 \\
        \midrule
        \multicolumn{1}{c}{2} & 1 & 1 & 0 & 0 \\
        \multicolumn{1}{c}{4} & 1 & 3 & 6 & 6 \\
        \multicolumn{1}{c}{9} & 1 & 8 & 56 & 336 \\
        \bottomrule
    \end{tabular}
    \caption{Number of paths from the primary case to a given susceptible.}
    \label{tab:paths}
\end{table}

The binomial variance assumes that infections in different household members are independent.
Because the each new infection in a household increases the risk of infection in the remaining susceptibles, infections within a household are positively correlated.
This correlation makes the true variance in the number of infections larger than the binomial variance.
To address this issue, cluster-adjusted variances~\cite{halloran2003estimating, cauchemez2009household, france2010household, carcione2011secondary, reichler2018household} and random effects~\cite{marsh2018unwelcome} have been used to account for correlation among household members.
Because of the bias in the point estimate of the SAR, this adjustment for clustering does not produce confidence intervals that have the expected coverage probabilities.

In a disease where the latent period (between infection and the onset of infectiousness) is longer than the infectious period, multiple generations of infection can be separated in time. 
%For example, measles has an average latent period of 10-12 days and an infectious period of approximately 8 days~\cite{heymann2008control}.
This was seen seen most famously by Peter Panum in a measles epidemic on the Faroe Islands in 1846~\cite{panum1940observations}.
With such separation, a binomial model could be used to estimate the risk of infection within a follow-up interval designed to capture only the first generation of transmission. 
However, most infectious diseases can have overlapping generations of infection.
For example, influenza has an average latent period of 1-2 days days and an infectious period of 3-4 days~\cite{longini2005containing}.
In general, the binomial model cannot be salvaged by adjusting the follow-up time of households.

In its original usage, the SAR was defined as the probability that a susceptible in a household with a primary case is infected by within-household transmission, whether or not there were multiple generations of transmission within the household~\cite{frost1938familial, wilson1939measles}. 
Here, we will call this the household final attack risk (FAR).
With complete follow-up of all households, a cluster-adjusted binomial model could produce an unbiased estimate of the FAR.
However, the estimated FAR will be biased upward if there are co-primary cases or if household members are at risk of infection from outside the household during the follow-up period~\cite{wilson1939measles, fox1974family, kemper1980error}.
Such conditions are common in practice, so the binomial model cannot be salvaged by returning to early interpretations of the SAR.

In its modern interpretation, the household SAR is an extremely useful measure of the transmissibility of infection.
However, this interpretation requires us to abandon the use of binomial models for estimation.
Here, we use probability generating functions and simulations to show that (1) a binomial model produces biased estimates of the household SAR even when the probability of transmission is small and (2) cluster adjustment of the variances does not produce interval estimates with the expected coverage probabilities.
To estimate the household SAR, explicit statistical models of disease transmission such as longitudinal chain binomial models~\cite{becker1989analysis, RampeyLongini1992} or pairwise survival analysis~\cite{kenah2011contact, kenah2013non, kenah2015semiparametric} should always be used.
We illustrate the practical implications of these results using household surveillance data collected by the Los Angeles County Department of Public Health during the 2009 influenza A (H1N1) pandemic.

\section{Methods}
For simplicity, our analytical calculations and simulations assume a uniform SAR within households (i.e., no variation in infectiousness or susceptibility) and no risk infection from outside the household except for the primary case.
These assumptions are not realistic: 
We intend to show that binomial models break down even under these ideal conditions.
We use probability generating functions (PGFs) to calculate the true outbreak size distributions at different combinations of the number of susceptibles ($m$) and the SAR ($p$), and we verify these calculations in simulations of household outbreaks. 

\subsection{Household outbreak size distributions}
Assume that each infectious member of a household makes infectious contact with each other member of the household with probability $p$ during his or her infectious period. 
Let $p_{mi}$ be the probability that $i$ out of $m$ susceptibles are infected by within-household transmission in a household with a single primary case. 
Then
\begin{equation}
    g_m(x) = \sum_{i = 0}^m p_{mi} x^i
    \label{eq:pgf}
\end{equation}
is the probability generating function (PGF) for the outbreak size distribution in a household with $m$ susceptibles and one primary case. 
Because a household with zero susceptibles has zero secondary infections with probability one, $g_0(x) = 1$.

The PGF for the outbreak size distribution in a household with $m + 1$ susceptibles can be derived from the PGFs for smaller households. 
Imagine a household with $m$ susceptibles of whom $i$ were infected. Now imagine that the household had one more susceptible. 
There are two possible outcomes:
\begin{enumerate}
  \item With probability $(1 - p)^{i + 1}$, the additional susceptible escapes infection from all $i + 1$ infected household members. The total number of infections in the household is $i$.
  \item With probability $1 - (1 - p)^{i + 1}$, the additional susceptible gets infected. He or she acts like a primary case in a household containing the $m - i$ susceptibles who escaped infection. There are $i + 1$ infections, and the number of infections among the remaining susceptibles has the PGF $g_{m - i}(x)$.
\end{enumerate}
Combining these results, we conclude that
\begin{equation}
    g_{m + 1}(x) = \sum_{i = 0}^m p_{mi} \big[(1 - p)^{i + 1} x^i + \big(1 - (1 - p)^{i + 1}\big) x^{i + 1} g_{m - i}(x)\big]
\end{equation}  
The first few iterations yield
\begin{align}
    g_0(x) &= 1, \\
    g_1(x) &= (1 - p) + p x, \\
    g_2(x) &= (1 - p)^2 + 2 p (1 - p)^2 x + (3p^2 - 2p^3) x^2 
\end{align}
which can be checked by hand. 
We calculated these polynomials using Python code in~\nameref{S2file}. 
As shown in Eq~\eqref{eq:pgf}, the coefficient on $x^i$ in the PGF $g_m(x)$ is the probability that $i$ of $m$ susceptibles are infected in a household outbreak started by a single primary case. 
Using these probabilities, we can calculate the mean and variance of the number of infections among the $m$ susceptibles. 

\subsection{Household outbreak simulations}
\label{sec:hhsims}
We simulated household outbreaks using Erd\H{o}s-R\'enyi random graphs~\cite{bollobas1998random, RN1}, where each pair of nodes is connected independently with probability $p$. 
In our graphs, each node represents a household member and $p$ is the SAR. One node is fixed as the primary case, and all household members connected to the primary case by a series of edges are infected.

We performed 40,000 simulations for each combination of household size and SAR. 
In each simulation, there were 200 independent households of the same size. 
We used logistic regression to calculate the proportion of susceptible household members who were infected with a naive 95\% confidence interval. 
We then calculated a cluster-adjusted confidence interval using generalized estimating equations (GEE) with a robust variance estimate.
The variance inflation factor (VIF) was calculated as the ratio of the robust variance to the naive variance.
All confidence intervals were calculated on the logit scale as $\hat{\beta} \pm 1.96\, \sigma$ where $\hat{\beta} = \logit(\hat{p})$ is the estimated log odds of infection and $\sigma$ is the naive or robust standard error estimate. 
Finally, we transformed the confidence intervals to the probability scale and estimated the coverage probabilities for the true household SAR and the true household FAR.

\paragraph{Source code}
Simulations were implemented in Python~3~\cite{Python3}, and statistical analysis was performed in~\texttt{R}~\cite{Rlang}. 
The \texttt{R} code is available in~\nameref{S1file}, and the Python code is available in~\nameref{S2file}.
All software used is free and open-source, and further details are given in the Supporting Information.

\subsection{Household data analysis}
\label{sec:LAdata}
To give a practical example of the consequences of using a binomial model to estimate the household SAR, we use influenza A (H1N1) household surveillance data collected by the Los Angeles County Department of Public Health (LACDPH) between April~22 and May~19, 2009. 
The data was collected using the following protocol~\cite{kenah2011contact}:
\begin{enumerate}
  \item Nasopharyngeal swabs and aspirates were taken from individuals who reported to the LACDPH or other health care providers with acute febrile respiratory illness (AFRI), defined as a fever $\geq 100^\circ \text{F}$ plus cough, core throat, or runny nose. 
    These specimens were tested for influenza, and the age, gender, and symptom onset date of the AFRI patient were recorded.
  \item Patients whose specimens tested positive for pandemic influenza A (H1N1) or for influenza A of undetermined subtype were enrolled as primary cases. 
    Each of them was given a structured phone interview to collect information about his or her household contacts. 
    They were asked to report the symptom onset date of any AFRI episodes among their household contacts. 
  \item When necessary, a follow-up interview was given 14 days after the symptom onset date of the primary case to assess whether any additional AFRI episodes had occurred in the household, including their illness onset date.
\end{enumerate}
For simplicity, we assume all AFRI episodes among household members were caused by influenza A (H1N1) and that all household members except the primary case were susceptible to infection. 
All analyses use natural history assumptions adapted from Ref~\cite{yang2009transmissibility} and identical to those in Refs~\cite{kenah2013non, kenah2015semiparametric}. 
In the primary analysis, we assumed an incubation period of 2~days, a latent period of zero~days, and an infectious period of 6~days. 
In a sensitivity analysis, we consider 7-day and 12-day infectious periods.

We estimated the household SAR for 2009 pandemic influenza A (H1N1) using binomial models, a longitudinal chain binomial model, and parametric pairwise regression models.
In each household, we censored observations at the end of the infectious period of the primary case.
Thus, the models are fit only to infections that could have been caused by primary cases, giving the binomial models the best possible chance of accurately estimating the household SAR.
For each assumed infectious period, all statistical models were fit to exactly the same data.
For simplicity, we did not include any covariates in these analyses.
Final size chain binomial models were not used because they require complete observation of each within-household epidemic, so they cannot be fit to data censored at the end of the infectious period of the primary case in each household.

\paragraph{Binomial models} 
Two binomial models were fit to the LACDPH households data. 
First, we used an intercept-only logistic regression model with unadjusted and cluster-adjusted confidence intervals~\cite{cameron2011robust}. 
Second, we used an intercept-only binomial GEE model~\cite{liang1986longitudinal} to get a second set of cluster-adjusted confidence intervals.

\paragraph{Longitudinal chain binomial model} The chain binomial model assumes that a given infectious person $A$ makes infectious contact with a given susceptible household member $B$ with an unknown probability $p$ on each day that $A$ is infectious.
On day $t$, an individual $B$ who is exposed to $k$ infectious household members will escape infection with probability $q^k$ and be infected with probability $1 - q^k$, where $q =  1 - p$.
The likelihood contribution from observation of individual $B$ is the product of these likelihood contributions over all days where $B$ was at risk of infection.
The overall likelihood is the product of the likelihood contributions of all susceptibles who were at risk of infection for at least one day. 

The household SAR is $1 - q^\iota$ where $\iota$ is the infectious period.
Because $p \in (0, 1)$, our likelihood was defined in terms of $\logit(p) = \ln (\sfrac{p}{q})$. 
To get a point estimate of the SAR, the unknown true $q$ is replaced by a point estimate $\hat{q} = 1 - \hat{p}$. 
Standard maximum likelihood estimation was used to get point and interval estimates on the logit scale, which were transformed back to the probability scale. 
For simplicity, we have assumed that the probability of escaping infection from an infectious household member does not depend on how long he or she has been infectious or on any covariates. 
More sophisticated longitudinal chain binomial models can allow the escape probability to vary with the time since infection or with covariates~\cite{becker1989analysis, RampeyLongini1992}. 

\paragraph{Pairwise survival analysis}
Pairwise survival analysis estimates failure times in ordered pairs consisting of an infectious individual and a susceptible household member~\cite{kenah2019handbook}. 
The pair AB is at risk of transmission starting with the onset of infectiousness in A, and failure occurs if A infects B. This failure time, called a \emph{contact interval} is right-censored if B is infected by someone other than A or if observation of the pair stops. 
To account for uncertainty about who-infected-whom, the overall likelihood is the sum of the likelihoods for all possible combinations of who-infected-whom consistent with the data~\cite{kenah2011contact}.
The survival function $S(\tau, \theta)$ is the probability that the contact interval is greater than $\tau$, where $\theta$ is a parameter vector. 
If $\theta_0$ is the true value of the parameter and the infectious period is $\iota$, then the household SAR is $1 - S(\iota, \theta_0)$. 
To get a point estimate of the SAR, the unknown true parameter $\theta_0$ is replaced by the maximum likelihood estimate $\hat{\theta}$. 

We used intercept-only exponential, Weibull, and log-logistic regression models~\cite{sharker2019pairwise}. 
For the exponential distribution, $S(\tau, \lambda) = \exp(-\lambda \tau)$ where $\lambda$ is the rate parameter. 
For the Weibull distribution, $S(\tau, \lambda, \gamma) = \exp[-(\lambda \tau)^\gamma]$ where $\lambda$ is the rate and $\gamma$ is the shape parameter. 
For the log-logistic distribution, $S(\tau, \lambda, \gamma) = [1 + (\lambda \tau)^\gamma]^{-1}$ for rate $\lambda$ and shape $\gamma$. 
For all three distributions, $\lambda > 0$ and $\gamma > 0$ so we defined our likelihoods in terms of their natural logarithms $\ln \lambda$ and $\ln \gamma$. 
Standard maximum likelihood estimation was used to get point estimates and a covariance matrix for the rate and shape parameters. 
To get a 95\% confidence interval for the SAR, we sampled $\ln \lambda$ and $\ln \gamma$ from their approximate multivariate normal distribution, calculated the household SAR for each sample, and took the $2.5\%$ and $97.5\%$ quantiles of the calculated $SAR$s as confidence limits.

\paragraph{Goodness of fit} 
To see how well the SAR estimates fit the data, we simulated outbreaks in the Los Angeles households using SAR point estimates from the binomial model, the chain binomial model, and pairwise survival models. 
In each simulation, we calculated the total number of infections among susceptible household members. 
For each SAR estimate, we performed 4,000 simulations. 
We then compared the simulated household epidemics to the observed final size of the outbreak started by the primary cases (i.e., the total number of cases who can be linked to a primary case through one or more generations of transmission). 
For all infectious periods shorter than 12 days, there are a few observed cases that occur after the end of the initial within-household outbreak.
Given the assumed infectious period, these late cases are excluded because they can only be explained by later introductions of infection into the household.

\paragraph{Source code} 
Statistical analyses were done with \texttt{R}~\cite{Rlang}, and the simulations were implemented in Python~3~\cite{Python3}.
The \texttt{R} code is available in~\nameref{S3file}, the Python code is available in~\nameref{S4file}, and the household data are available in~\nameref{S5file}.
All software used is free and open-source, and further details are given in the Supporting Information.

\section{Results}
\subsection{Household outbreak simulations}
\begin{figure}
    \centering
    \includegraphics[width = \textwidth]{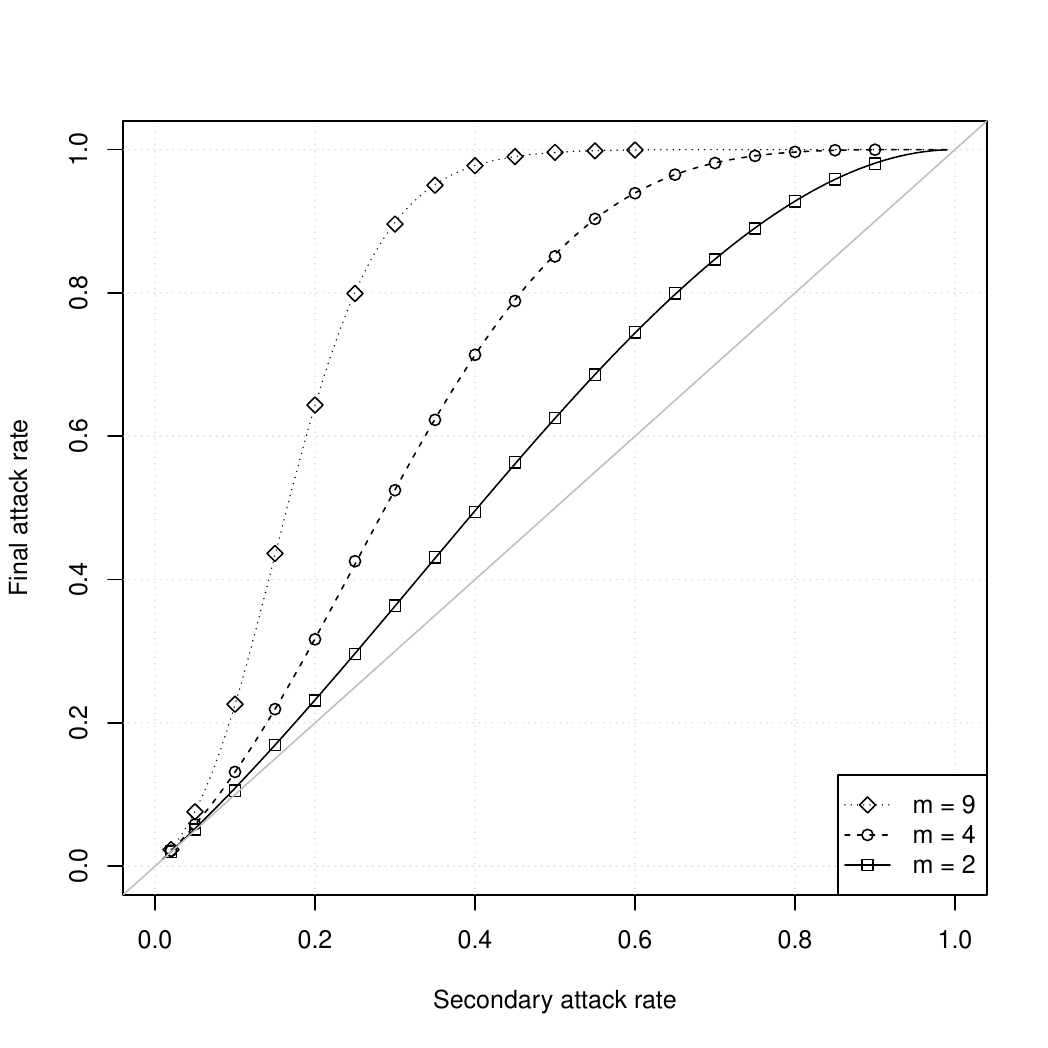}
    \caption{The household FAR as a function of the SAR for households with different numbers of susceptibles $m$.
    Lines show analytical calculations using probability generating functions, and simulations show estimates from 40,000 simulated household outbreaks.
    Each simulated household outbreak had a single primary case, so the total household size was $m + 1$.}
    \label{fig:far}
\end{figure}

Fig~\ref{fig:far} shows the household FAR calculated using PGFs (lines) and from simulations (symbols) as a function of the true SAR and the number of susceptibles. 
There is excellent agreement between the analytical calculations and the simulations. 
Both show that the household FAR is larger than the household SAR when there is more than one susceptible. 
At a fixed SAR, the difference between the SAR and the FAR increases with household size. 
Thus, a binomial model will produce a point estimate of the SAR that is biased upward whenever there is more than one susceptible household member. 

\begin{figure}
    \centering
    \includegraphics[width = \textwidth]{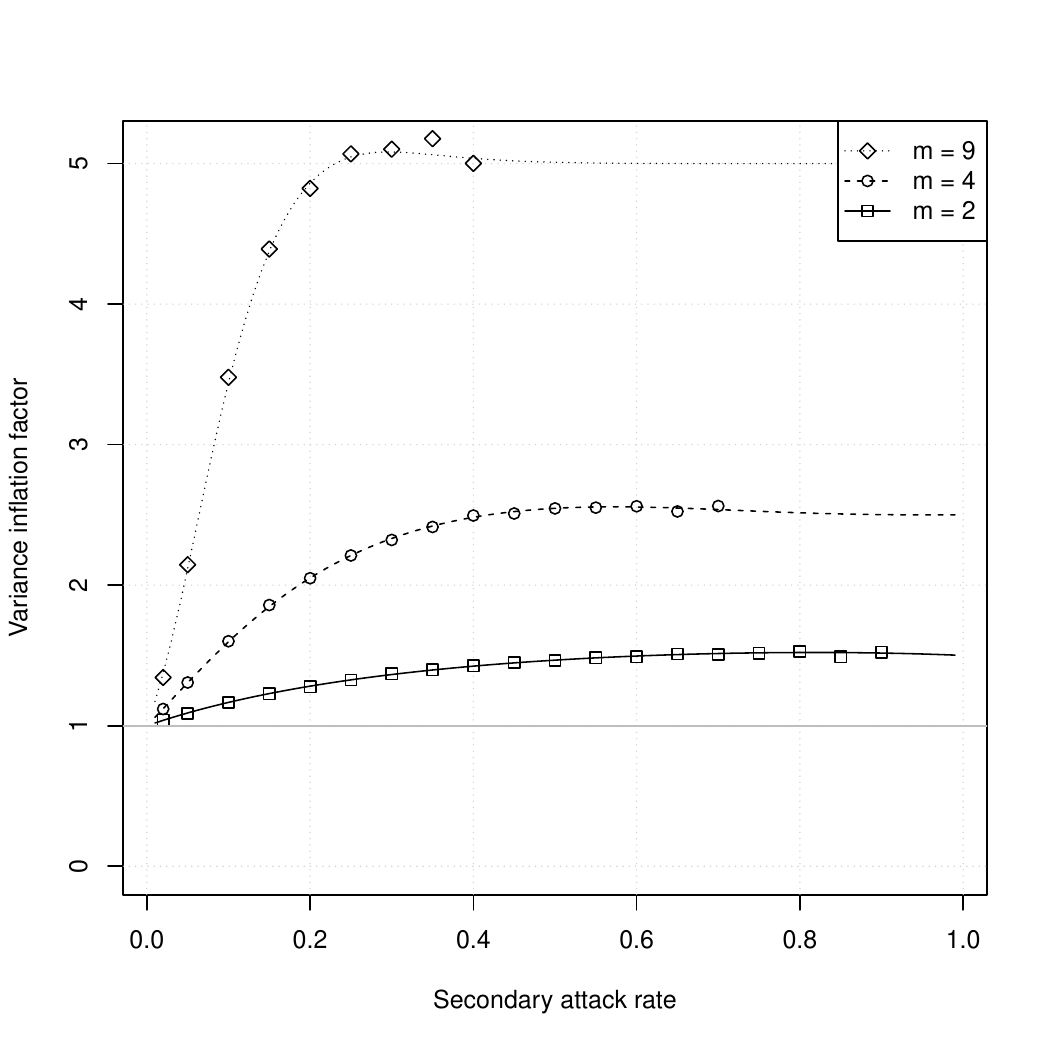}
    \caption{The VIF as a function of the SAR for households with $m$ susceptibles. 
    Lines show analytical calculations, and symbols show estimates from 40,000 simulated household outbreaks.
    Each simulated household outbreak started with a single primary case, so the total household size was $m + 1$.
    For numerical stability, symbols are shown only for simulations with an observed FAR $< 0.99$.}
    \label{fig:vif}
\end{figure}

Fig~\ref{fig:vif} shows the VIF calculated using PGFs (lines) and from simulations (symbols) as a function of the true SAR and the number of susceptibles.
Again, there is excellent agreement between the analytical calculations and the simulations.
The variance of the number of infections within households is substantially larger than the binomial variance, and this difference increases with increasing household size.
Thus, confidence intervals based on a binomial estimate will have coverage probabilities that are too low even if the estimated SAR is correct.

\begin{figure}
    \centering
    \includegraphics[width = \textwidth]{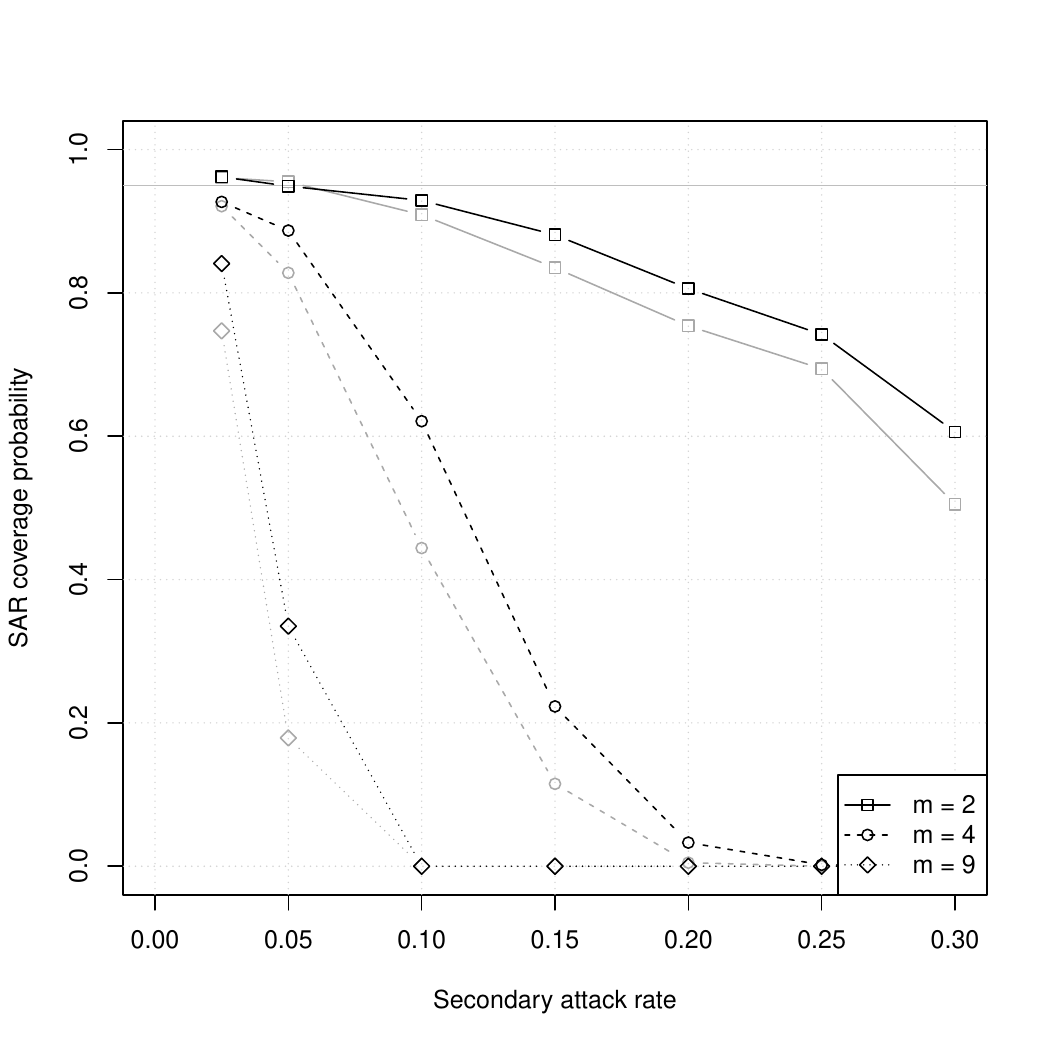}
    \caption{Coverage probabilities of binomial 95\% confidence intervals for the household SAR with different numbers of susceptibles ($m$). 
    Gray lines are coverage probabilities for unadjusted confidence intervals, and black lines are coverage probabilities for cluster-adjusted confidence intervals. 
    Each symbol represents 1,000 simulations with 100 households each.}
    \label{fig:ci_SAR}
\end{figure}

Fig~\ref{fig:ci_SAR} shows the household SAR coverage probabilities for unadjusted and cluster-adjusted binomial 95\% confidence intervals. 
Even for small households, the coverage probabilities are below 95\% and decrease rapidly as the true SAR increases. 
Cluster adjustment increases the coverage probabilities only slightly. 
With or without adjustment for clustering by household, a binomial model does not produce reliable point or interval estimates of the household SAR.

\begin{figure}
    \centering
    \includegraphics[width = \textwidth]{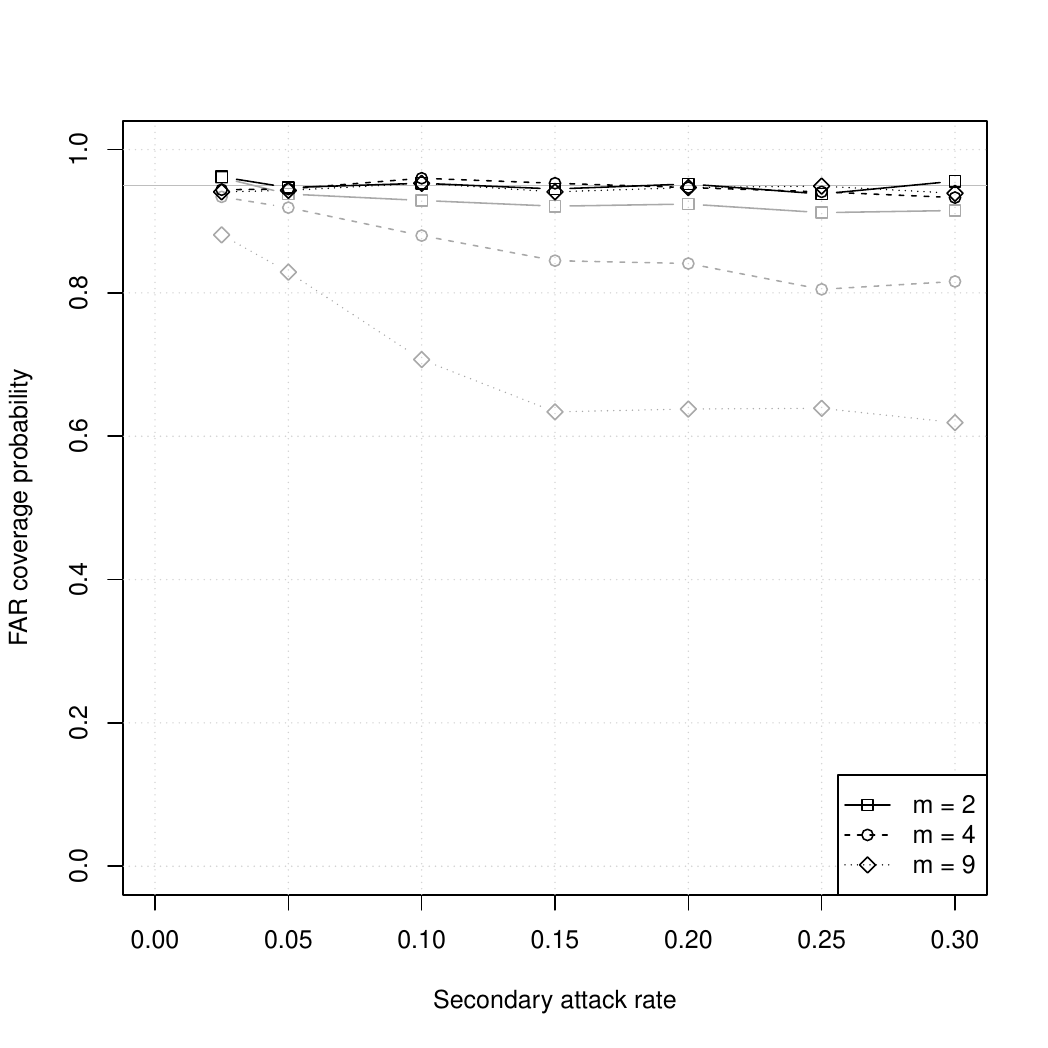}
    \caption{Coverage probabilities of binomial 95\% confidence intervals for the household FAR with different numbers of susceptibles ($m$). 
    Gray lines are coverage probabilities for unadjusted confidence intervals, and black lines are coverage probabilities for cluster-adjusted confidence intervals.
    Each symbol represents 1,000 simulations with 100 households each.}
    \label{fig:ci_FAR}
\end{figure}

Fig~\ref{fig:ci_FAR} shows coverage probabilities of unadjusted and cluster-adjusted 95\% confidence intervals for the household FAR.  
Coverage of the FAR is much higher than coverage of the SAR.
However, the coverage probabilities for unadjusted confidence intervals are always below 95\%, and they decrease with increasing household size or increasing SAR. 
Adjustment for clustering by household corrects this problem, producing coverage probabilities close to 95\% for all household sizes. 
Under these ideal conditions, a binomial model can produce reliable point and interval estimates of the household FAR as long as clustering within households is taken into account.
This does not imply that FAR can be defined clearly or estimated accurately under more realistic conditions, and it does not imply that the FAR is an acceptable substitute for the SAR in practice.

\subsection{Household data analysis}
In the LACDPH pandemic influenza A (H1N1) data, there were 58 households with a total of 299 members. 
There were 99 infections, of which 62 were classified as primary cases because 4 of 58 households had two co-primary cases.
There were 37 household contacts who were infected while under observation.
The median household size was 5 with a range from 2 to 20.
Both in this example and more generally, co-primary cases and varying household sizes are practical problems for estimation of the household SAR. 

There are three types of cases relevant to our analyses:
\emph{Possible second generation cases} are susceptible household members who are infected during the infectious period of the primary case, so it is possible that they were infected by the primary case.
\emph{Final size cases} are susceptible household members who could have been infected through a chain of transmission starting from a primary case.
\emph{Late cases} are susceptible household members who were infected after the end of the infectious period of the last final size case in the household. 
Given the assumed infectious period, these cases can only be explained by a new introduction of infection to the household. 
They are excluded from SAR estimation and from the observed final size for the assumed infectious period.

Table~\ref{tab:cases} shows the numbers of possible second generation cases, final size cases, and late cases for each assumed infectious period from 3 days (almost certainly too short) to 12~days (almost certainly too long).
Assuming an infectious period of 6~days, there are 24 possible second generation cases, 26~final size cases, and 11~late cases.
Assuming an infectious period of 7~days results in substantially larger numbers of possible second generation cases and final size cases.
An infectious period of at least 12~days is required to account for all observed cases through within-household transmission.
We show analyses with 6-day, 7-day, and 12-day infectious periods.

\begin{table}
  \centering
  \begin{tabular}{cccc}
    \toprule
    Infectious period & Possible second   & Final size  & Late \\
    (days)            &  generation cases & cases       & cases \\
    \midrule
    3   & 13  & 16  & 21 \\
    4   & 17  & 22  & 15 \\
    5   & 20  & 25  & 12 \\
    6   & 24  & 26  & 11 \\
    7   & 28  & 32  & 5 \\
    8   & 28  & 32  & 5 \\
    9   & 28  & 32  & 5 \\
    10  & 32  & 36  & 1 \\
    11  & 33  & 36  & 1 \\
    12  & 34  & 37  & 0 \\
    \bottomrule
  \end{tabular}
  \caption{The number of possible second generation cases, final size cases, and late cases for each assumed infectious period. There are always 37 total final size and late cases.}
  \label{tab:cases}
\end{table}

Table~\ref{tab:est} shows point estimates and 95\% confidence intervals for the household SAR. 
The point estimates for all binomial models are identical.
As expected, binomial models produce much higher estimates than the chain binomial or pairwise regression models.
Adjustment for clustering produced a wider confidence interval, with cluster-adjusted variance and GEE producing very similar results. 
The chain binomial and exponential pairwise regression models produced nearly identical point and interval estimates of the household SAR. 
For each infectious period, the Weibull and log-logistic pairwise regression models produced slightly different SAR estimates and wider confidence intervals than the exponential model.
In all cases, the exponential model had the lowest AIC.
The chain binomial and pairwise regression estimates are consistent with each other, but neither is consistent with the binomial estimates.

\begin{table}
  \centering
  \begin{tabular}{llccc}
    \toprule
    \multicolumn{2}{c}{Model}   & \multicolumn{2}{c}{Estimated SAR} & AIC\\
    \midrule
    \multicolumn{5}{l}{\textbf{6-day infectious period}}\\
    Binomial: & GLM (naive)               & 0.101 & (0.067, 0.144) &\\
              & GLM (adjusted)            & 0.101 & (0.052, 0.189) &\\
              & GEE (naive)               & 0.101 & (0.069, 0.147) &\\
              & GEE (robust)              & 0.101 & (0.052, 0.188) &\\[5pt]
    \multicolumn{2}{l}{Longitudinal chain binomial} 
                                          & 0.076 & (0.051, 0.109) &\\[5pt]
    Pairwise regression:  & exponential   & 0.075 & (0.051, 0.110) & 235.96\\
                          & Weibull	  & 0.079 & (0.056, 0.138) & 236.38\\
                          & log-logistic  & 0.079 & (0.055, 0.133) & 236.26\\
    \midrule
    \multicolumn{5}{l}{\textbf{7-day infectious period}} \\
    Binomial: & GLM (naive)               & 0.118 & (0.081, 0.163) &\\
              & GLM (adjusted)            & 0.118 & (0.063, 0.211) &\\
              & GEE (naive)               & 0.118 & (0.083, 0.166) &\\
              & GEE (robust)              & 0.118 & (0.063, 0.210) &\\[5pt]
    \multicolumn{2}{l}{Longitudinal chain binomial}
                                          & 0.087 & (0.059, 0.121) &\\[5pt]
    Pairwise regression:  & exponential   & 0.086 & (0.060, 0.122) & 273.20\\
                          & Weibull	  & 0.089 & (0.064, 0.146) & 274.12\\
                          & log-logistic  & 0.089 & (0.064, 0.144) & 273.97\\
    \midrule
    \multicolumn{5}{l}{\textbf{12-day infectious period}}\\
    Binomial: & GLM (naive)               & 0.143 & (0.103, 0.192) &\\
              & GLM (adjusted)            & 0.143 & (0.084, 0.234) &\\
              & GEE (naive)               & 0.143 & (0.104, 0.194) &\\
              & GEE (robust)              & 0.143 & (0.085, 0.233) &\\[5pt]
    \multicolumn{2}{l}{Longitudinal chain binomial}
                                          & 0.097 & (0.069, 0.131) &\\[5pt]
    Pairwise regression:  & exponential   & 0.097 & (0.070, 0.134) & 362.91\\
                          & Weibull	  & 0.095 & (0.071, 0.148) & 364.33\\
                          & log-logistic  & 0.095 & (0.070, 0.144) & 363.99\\
    \bottomrule
  \end{tabular}
  \caption{Estimates of the household SAR with 95\% confidence limits and Akaike information criterion (AIC) for pairwise regression models.}
  \label{tab:est}
\end{table}

\begin{figure}
    \centering
    \includegraphics[width = \textwidth]{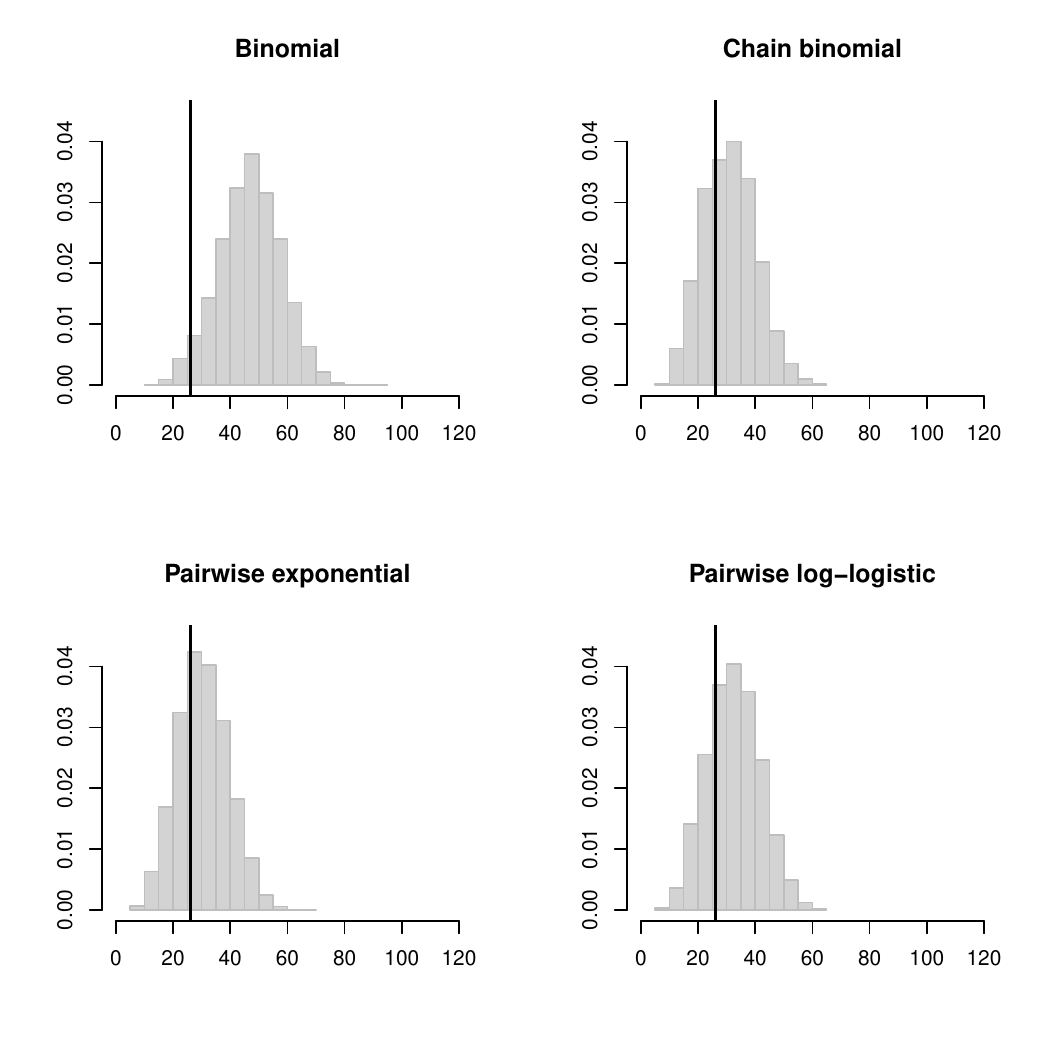}
    \caption{Histograms of simulated final outbreak sizes in the LA households based on household SAR estimates assuming a 6-day infectious period. Vertical black lines indicate the observed final size of 26 cases.}
    \label{fig:hist6}
\end{figure}

\begin{figure}
    \centering
    \includegraphics[width = \textwidth]{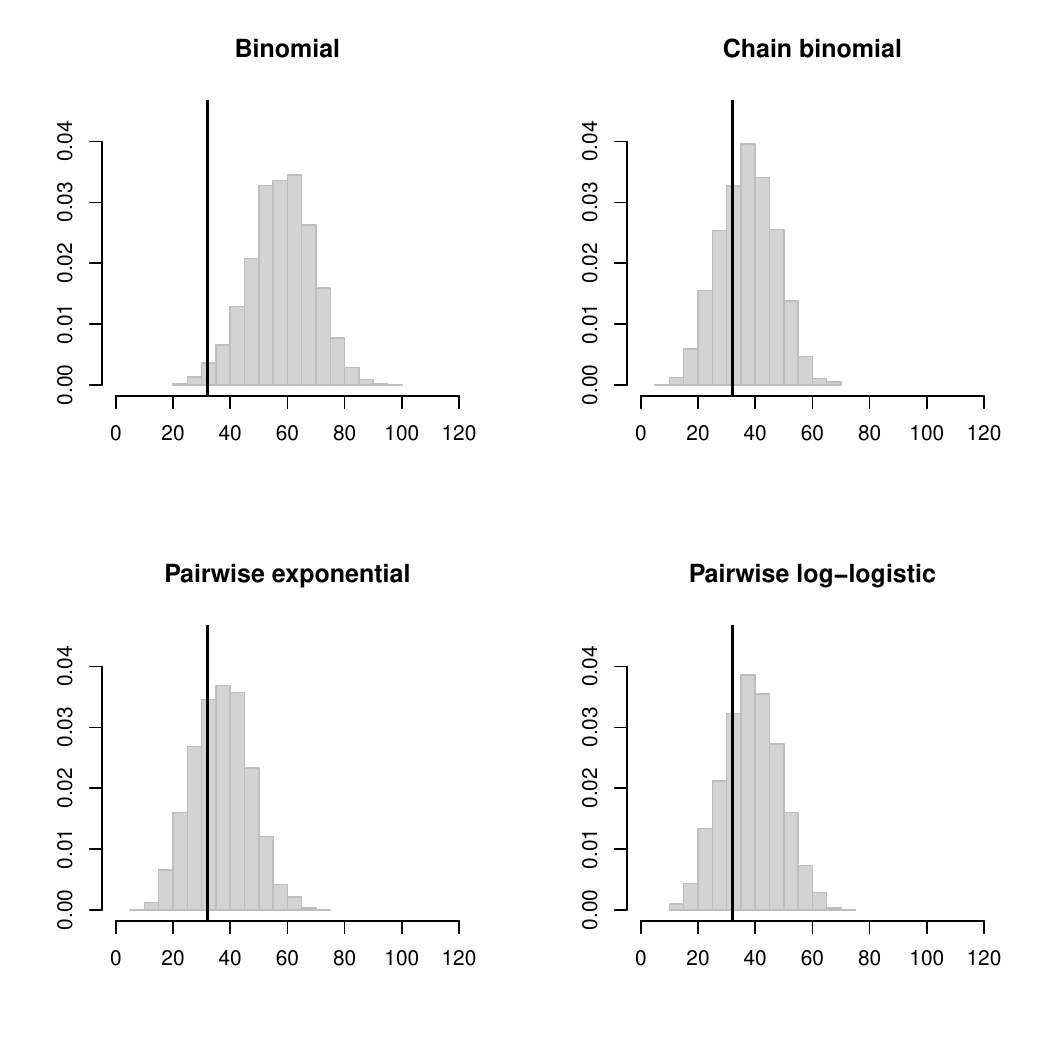}
    \caption{Histogram of simulated final outbreak sizes in the LA households based on SAR estimates assuming a 7-day incubation period. Vertical black lines indicate the observed final size of 32 cases.}
    \label{fig:hist7}
\end{figure}

\begin{figure}
    \centering
    \includegraphics[width = \textwidth]{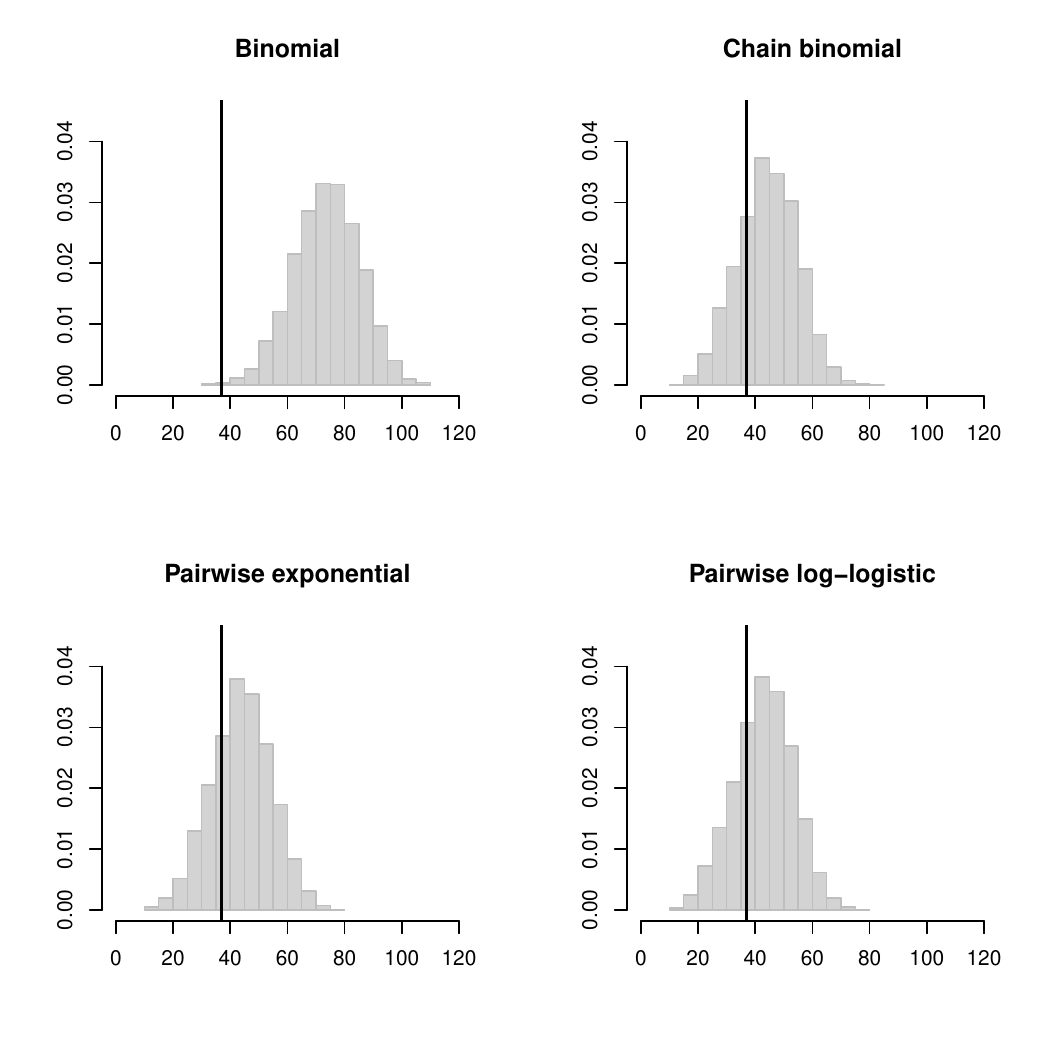}
    \caption{Histogram of simulated final outbreak sizes in the LA households based on SAR estimates assuming a 12-day infectious period. Vertical black lines indicate the observed final size of 37 cases.}
    \label{fig:hist12}
\end{figure}

Fig~\ref{fig:hist6} shows histograms of the simulated outbreak sizes in the LA households based on the four different SAR estimates that assume a  6-day infectious period. 
The binomial estimates predict outbreaks larger than observed, but the chain binomial and pairwise estimates predict outbreak size distributions centered near observed outbreak size.
Fig~\ref{fig:hist7} and Fig~\ref{fig:hist12} shows a similar pattern for estimates that assume 7-day and 12-day infectious periods, respectively.
For the binomial estimates, the predicted outbreak sizes increase quickly with the assumed infectious period.
For the chain binomial and pairwise regression estimates, the predicted outbreak sizes increase much more slowly.
To the extent that a true household SAR exists, it is almost certainly below the binomial estimates and closer to the chain binomial and pairwise regression estimates.

\begin{figure}
    \centering
    \includegraphics[width = \textwidth]{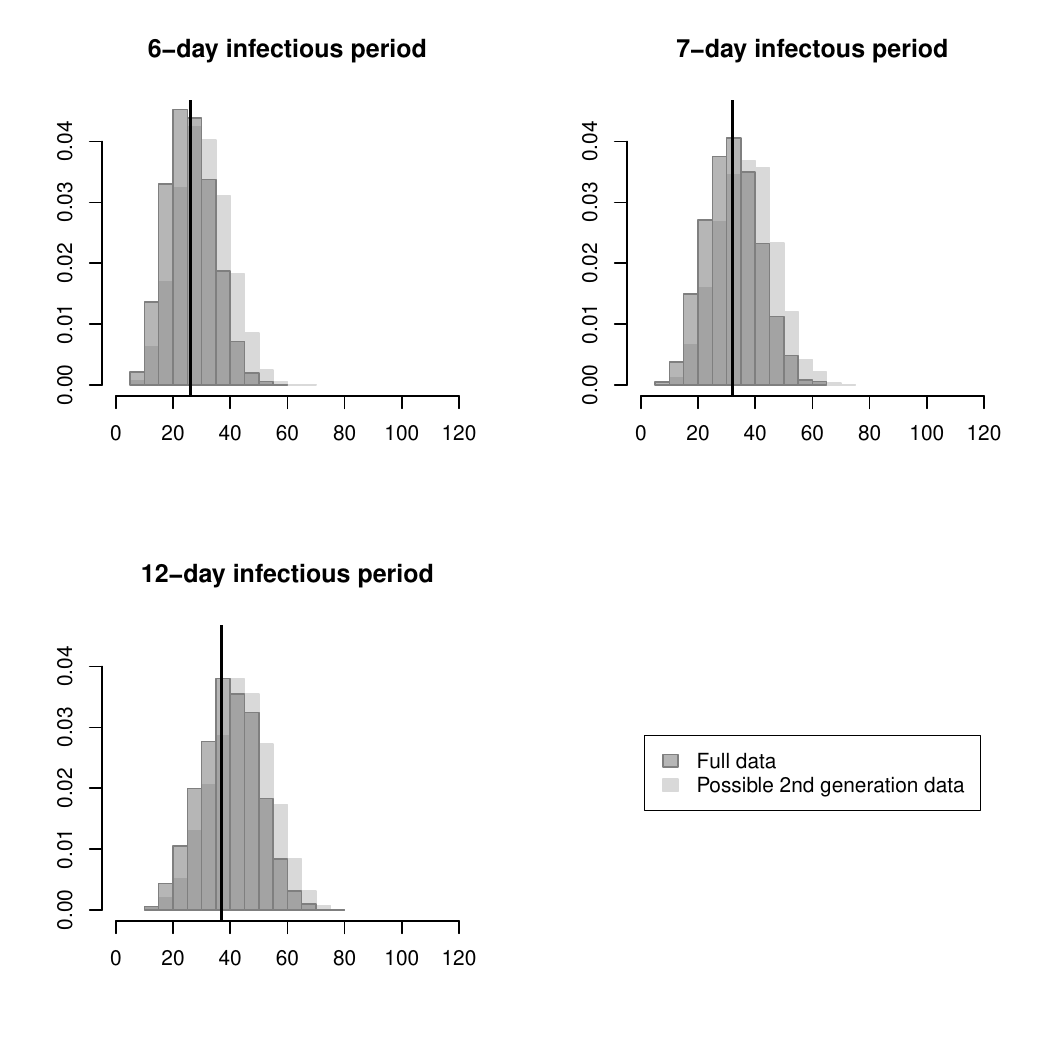}
    \caption{Histograms of simulated outbreak sizes based on pairwise exponential SAR estimates using the full data (dark gray) superimposed on the corresponding histograms from Figs.~\ref{fig:hist6}--\ref{fig:hist12} based on estimates using second generation data (light gray). 
    For each assumed infectious period, a vertical black line shows the observed final outbreak size.}
    \label{fig:fullhist}
\end{figure}

An important advantage of the longitudinal chain binomial and pairwise regression models is that they can estimate the SAR using the entire period of household observation even when there is an ongoing risk of infection from outside the household.
Table~\ref{tab:fullest} shows point and interval estimates of the SAR based on the full data set collected by the LACDPH. 
As before, the chain binomial and pairwise exponential models produce nearly identical point and interval estimates.
Using the full data set, the pairwise Weibull and log-logistic models produce point estimates closer to those of the one-parameter models than in Table~\ref{tab:est}, but their confidence intervals remain wider.
All four models produce lower point estimates of the SAR when using the full data set than when using only the possible second generation data. 
Fig~\ref{fig:fullhist} shows the distribution of outbreak sizes under the pairwise exponential estimate of the SAR assuming 6-, 7-, and 12-day infectious periods. 
The light gray histograms in the background show the distributions based on the point estimates from Table~\ref{tab:est}, which used the possible second generation data.
In all three cases, there is a small but clear improvement in the predictive fit of the model when the full data set is used. 
Similar results were seen for the longitudinal chain binomial and pairwise Weibull and logistic regression models (not shown but produced by~\nameref{S3file}). 

\begin{table}
  \centering
  \begin{tabular}{llccc}
    \toprule
    \multicolumn{2}{c}{Model}   & \multicolumn{2}{c}{Estimated SAR} & AIC\\
    \midrule
    \multicolumn{5}{l}{\textbf{6-day infectious period}}\\
    \multicolumn{2}{l}{Longitudinal chain binomial} 
                                          & 0.069 & (0.048, 0.096) &\\[5pt]
    Pairwise regression:  & exponential   & 0.068 & (0.048, 0.097) & 288.33\\
                          & Weibull	  & 0.069 & (0.050, 0.111) & 289.74\\
                          & log-logistic  & 0.069 & (0.050, 0.111) & 289.54\\
    \midrule
    \multicolumn{5}{l}{\textbf{7-day infectious period}} \\
    \multicolumn{2}{l}{Longitudinal chain binomial}
                                          & 0.079 & (0.056, 0.108) &\\[5pt]
    Pairwise regression:  & exponential   & 0.078 & (0.056, 0.107) & 331.78\\
                          & Weibull	  & 0.079 & (0.058, 0.123) & 333.19\\
                          & log-logistic  & 0.079 & (0.058, 0.121) & 333.01\\
    \midrule
    \multicolumn{5}{l}{\textbf{12-day infectious period}}\\
    \multicolumn{2}{l}{Longitudinal chain binomial}
                                          & 0.091 & (0.066, 0.121) &\\[5pt]
    Pairwise regression:  & exponential   & 0.090 & (0.066, 0.124) & 399.34\\
                          & Weibull	  & 0.089 & (0.067, 0.131) & 399.75\\
                          & log-logistic  & 0.090 & (0.067, 0.129) & 399.34\\
    \bottomrule
  \end{tabular}
  \caption{Full-data estimates of the household SAR with 95\% confidence limits and Akaike information criterion (AIC) for pairwise regression models.}
  \label{tab:fullest}
\end{table}

\section{Discussion}
Studies of disease transmission in households and other clearly-defined groups at risk of infection are part of a glorious tradition in infectious disease epidemiology~\cite{frost1938familial}.
They remain one of the most effective means of obtaining critical information about routes and risk factors of transmission, the basic reproduction number, and the natural history of epidemic diseases~\cite{fox1974family, lipsitch2020defining}.
Every author of the studies cited above has made an important contribution to infectious disease epidemiology and to public health.
However, these studies should no longer be analyzed using binomial models.
Even when the SAR is small, it is important to account for multiple generations of transmission within households.
Unless these generations are clearly separated in time, a binomial estimate of the SAR will be biased upward and have a confidence interval with low coverage probability even if the variance is adjusted for clustering.

A binomial model can estimate the household FAR accurately if cluster-adjusted confidence intervals are used.
However, the FAR was clearly defined in our simulations only because we made the following assumptions: (1) Each household had at most one primary case, (2) susceptibles were not at risk of infection from outside the household, and (3) all households had the same size.
In practice, these assumptions are extremely unlikely to hold.
The LACDPH data had households with multiple primary cases and household sizes that varied from 2 to 20.
For all assumed infectious periods shorter than 12 days, there were cases that could only be explained by the re-introduction of infection to the household.
Unlike the FAR, the household SAR can be clearly defined and estimated even when there are multiple primary cases, ongoing risk of infection from outside the household, and varying household sizes.

The discrete-time chain binomial model~\cite{RampeyLongini1992} and pairwise survival models~\cite{kenah2011contact, kenah2013non, kenah2015semiparametric} require more detailed follow-up of each household than final size models, but they can account accurately for delayed entry, loss to follow-up, and the risk of infection from outside the household.
If there are asymptomatic infections or if infection times cannot be determined with sufficient precision, data augmentation and Markov chain Monte Carlo (MCMC) can be used to account for the transmission of infection~\cite{o1999bayesian}.
These longitudinal models also allow the probability or hazard of transmission to depend on individual-level, pairwise, and household-level covariates~\cite{kenah2015semiparametric}.
The household members in the LACDPH data varied in ways that could have affected their susceptibility and infectiousness, including age, sex, and use of antiviral prophylaxis.
Simultaneous estimation of these effects is  critical to preventing bias for contagious outcomes~\cite{morozova2018risk}. 
Accurate estimates of covariate effects on infectiousness and susceptibility can provide critical insight into the effectiveness of public health interventions such as handwashing, social distancing, antiviral prophylaxis or treatment, and vaccination.

Whereas binomial models can be fit using almost any standard statistical package, the lack of available software has been a major obstacle to the adoption of statistical models of infectious disease transmission in household studies.
Chain binomial models are available in the free and open source software package TranStat (\texttt{www.cidid.org/transtat}), which incorporates several advanced methods~\cite{Yang-resampling-2007, Yang-EM-2012} and has been used in analyses of influenza~\cite{yang2009transmissibility}, Zika virus~\cite{rojas2016epidemiology}, and Ebola virus~\cite{fang2016transmission}. 
Pairwise survival models  are available in the free and open source \texttt{transtat} package for \texttt{R}, which was used to analyze the LA household data above.
This package includes parametric models and semiparametric models~\cite{kenah2011contact, kenah2013non, kenah2015semiparametric, kenah2019handbook}.

In the COVID-19 pandemic, there have been too few studies of SARS-CoV-2 transmission in households or other clearly-defined populations at risk of infection, leaving unanswered many questions about the modes and intensity of transmission and the predictors of infectiousness and susceptibility~\cite{lipsitch2020defining}.
This has forced public health decisions that affect millions of lives to be made under crushing uncertainty.
Household studies can provide critical scientific insights to guide public health interventions and policies.
The results above show that replacing binomial models with statistical models of transmission will help infectious disease epidemiologists who conduct these studies contribute more effectively to the prevention and control of epidemics.

\section*{Supporting information}
% Include only the SI item label in the paragraph heading. 
% Use the \nameref{label} command to cite SI items in the text.
\paragraph*{S1 File.}
\label{S1file}
{\bf SARsims.R.} 
\texttt{R}~\cite{Rlang} (\url{https://www.r-project.org/}) code used to analyze the household outbreak simulations in Section~\ref{sec:hhsims}.
Produces Figs~\ref{fig:far} to \ref{fig:ci_FAR}.
Requires the following packages:
\begin{itemize}
  \item gee~\cite{Rgee} (\url{https://cran.r-project.org/package=gee})
  \item reticulate~\cite{Rreticulate} (\url{https://rstudio.github.io/reticulate/})
  \item survival~\cite{Rsurvival} (\url{https://cran.r-project.org/package=survival})
  \item transtat~\cite{Rtranstat} (\url{https://github.com/ekenah/transtat})
\end{itemize}
Directions and package versions used for publication are in comments.

\paragraph*{S2 File.}
\label{S2file}
{\bf SARsims.py.} 
Python 3~\cite{Python3} (\url{https://www.python.org}) functions called by~\nameref{S1file}.
Requires the following packages:
\begin{itemize}
  \item NetworkX~\cite{hagberg2008exploring} (\url{https://networkx.github.io})
  \item NumPy~\cite{oliphant2006guide} and pandas~\cite{mckinney-proc-scipy-2010} (\url{https://www.scipy.org})
\end{itemize}
Directions and package versions used for publication are in comments.

\paragraph*{S3 File.}
\label{S3file}
{\bf LAanalysis.R.} 
\texttt{R}~\cite{Rlang} code used to analyze LACDPH household surveillance data in Section~\ref{sec:LAdata}.
Produces Tabs~\ref{tab:cases} to \ref{tab:fullest} and Figs~\ref{fig:hist6} to \ref{fig:fullhist}.
In addition to the packages listed in~\nameref{S1file}, the following packages are required:
\begin{itemize}
  \item MASS~\cite{Rmass} (\url{https://cran.r-project.org/package=MASS})
  \item sandwich~\cite{Rsandwich} (\url{https://cran.r-project.org/package=sandwich})
  \item stats4 (\url{https://cran.r-project.org/package=stats4})
\end{itemize}
Directions and package versions used for publication are in comments.

\paragraph*{S4 File.}
\label{S4file}
{\bf LAsims.py.} 
Python 3\cite{Python3} functions called by~\nameref{S3file}.
Requires the NetworkX and pandas packages listed under~\nameref{S2file}:
Directions and package versions used for publication are in comments.

\paragraph*{S5 File.}
\label{S5file}
{\bf LAdata\_2019-12.csv.} 
De-identified LACDPH household surveillance data in CSV format used by~\nameref{S3file}.

\section*{Acknowledgments}
The subtitle of the paper was inspired by Edsger Dijkstra's letter ``Go To Statement Considered Harmful'' (\emph{Communications of the ACM} 11:147--148, 1968). 
Brit Oiulfstad, Dee Ann Bagwell, Brandon Dean, Laurene Mascola, and Elizabeth Bancroft of the Los Angeles County Department of Public Health (LACDPH) generously provided the household influenza surveillance data. 

\paragraph{Funding}
EK and YS were supported by National Institute of Allergy and Infectious Diseases (NIAID) grants R01 AI116770 and R03 AI124017. 
EK was also supported by National Institute of General Medical Sciences (NIGMS) grant U54 GM111274, and YS was also supported by National Institutes of Health (NIH) grant DP2HD09179.
The funders had no role in study design, data collection and analysis, decision to publish, or preparation of the manuscript.

\paragraph{Disclaimer}
The contribution of YS was completed prior to his Food and Drug Administration (FDA) employment. 
The content is solely the responsibility of the authors and does not represent the official views or policies of LACDPH, NIAID, NIGMS, NIH, or FDA.

\bibliographystyle{plos2015}
\bibliography{SARestimation} 

\begin{thebibliography}{10}

\bibitem{morgenstern1980measures}
Morgenstern H, Kleinbaum DG, Kupper LL.
\newblock Measures of disease incidence used in epidemiologic research.
\newblock International Journal of Epidemiology. 1980;9(1):97--104.

\bibitem{de1981meningococcal}
De~Wals P, Hertoghe L, Borl\'{e}e-Grim\'{e}e I, De~Maeyer-Cleempoel S,
  Reginster-Haneuse G, Dachy A, et~al.
\newblock Meningococcal disease in {B}elgium. {S}econdary attack rate among
  household, day-care nursery and pre-elementary school contacts.
\newblock Journal of Infection. 1981;3:53--61.

\bibitem{fox1974family}
Fox JP.
\newblock Family-based epidemiologic studies.
\newblock American Journal of Epidemiology. 1974;99(3):165--79.

\bibitem{elveback1976influenza}
Elveback LR, Fox JP, Ackerman E, Langworthy A, Boyd M, Gatewood L.
\newblock An influmza simulation model for immunization studies.
\newblock American Journal of Epidemiology. 1976;103(2):152--165.

\bibitem{monto1994studies}
Monto AS.
\newblock Studies of the community and family: acute respiratory illness and
  infection.
\newblock Epidemiologic Reviews. 1994;16(2):351.

\bibitem{halloran2003estimating}
Halloran ME, Pr{\'e}ziosi MP, Chu H.
\newblock Estimating vaccine efficacy from secondary attack rates.
\newblock Journal of the American Statistical Association. 2003;98(461):38--46.

\bibitem{terry2014field}
Terry J, Flatley C, van~den Berg D, Morgan G, Trent M, Turahui J, et~al.
\newblock A field study of household attack rates and the effectiveness of
  macrolide antibiotics in reducing household transmission of pertussis.
\newblock Communicable Diseases Intelligence Quarterly Report.
  2014;39(1):E27--33.

\bibitem{frost1938familial}
Frost WH.
\newblock The familial aggregation of infectious diseases.
\newblock American Journal of Public Health and the Nations Health.
  1938;28(1):7--13.

\bibitem{wilson1939measles}
Wilson EB, Bennett C, Allen M, Worcester J.
\newblock Measles and scarlet fever in Providence, RI, 1929-1934 with respect
  to age and size of family.
\newblock Proceedings of the American Philosophical Society. 1939; p. 357--476.

\bibitem{terris1999charles}
Terris M.
\newblock Charles V. Chapin (1856-1941),``Dean of City Health Officers".
\newblock Journal of Public Health Policy. 1999;20(2):215--220.

\bibitem{jordan1958study}
Jordan~Jr WS, Denny~Jr FW, Badger GF, Dingle JH, Oseasohn R, Stevens D, et~al.
\newblock A study of illness in a group of Cleveland families. XVII. The
  occurrence of Asian influenza.
\newblock American Journal of Hygiene. 1958;68(2):190--212.

\bibitem{chin1960morbidity}
Chin TD, Foley JF, Doto IL, Gravelle CR, Weston J.
\newblock Morbidity and mortality characteristics of Asian strain influenza.
\newblock Public Health Reports. 1960;75(2):149.

\bibitem{davis1970hong}
Davis LE, Caldwell GG, Lynch RE, Bailey RE, Chin TD.
\newblock Hong Kong influenza: the epidemiologic features of a high school
  family study analyzed and compared with a similar study during the 1957 Asian
  influenza epidemic.
\newblock American Journal of Epidemiology. 1970;92(4):240--247.

\bibitem{goh2004secondary}
Goh DLM, Lee BW, Chia KS, Heng BH, Chen M, Ma S, et~al.
\newblock Secondary household transmission of SARS, Singapore.
\newblock Emerging infectious diseases. 2004;10(2):232.

\bibitem{cauchemez2004bayesian}
Cauchemez S, Carrat F, Viboud C, Valleron A, Boelle P.
\newblock A Bayesian MCMC approach to study transmission of influenza:
  application to household longitudinal data.
\newblock Statistics in Medicine. 2004;23(22):3469--3487.

\bibitem{tsang2014association}
Tsang TK, Cauchemez S, Perera RA, Freeman G, Fang VJ, Ip DK, et~al.
\newblock Association between antibody titers and protection against influenza
  virus infection within households.
\newblock The Journal of Infectious Diseases. 2014;210(5):684--692.

\bibitem{tsang2015influenza}
Tsang TK, Cowling BJ, Fang VJ, Chan KH, Ip DK, Leung GM, et~al.
\newblock Influenza A virus shedding and infectivity in households.
\newblock The Journal of Infectious Diseases. 2015;212(9):1420--1428.

\bibitem{petrie2017application}
Petrie JG, Eisenberg MC, Ng S, Malosh RE, Lee KH, Ohmit SE, et~al.
\newblock Application of an individual-based transmission hazard model for
  estimation of influenza vaccine effectiveness in a household cohort.
\newblock American Journal of Epidemiology. 2017;186(12):1380--1388.

\bibitem{banerjee2008evidence}
Banerjee I, Primrose~Gladstone B, Iturriza-Gomara M, Gray JJ, Brown DW, Kang G.
\newblock Evidence of intrafamilial transmission of rotavirus in a birth cohort
  in South India.
\newblock Journal of Medical Virology. 2008;80(10):1858--1863.

\bibitem{yang2009transmissibility}
Yang Y, Sugimoto JD, Halloran ME, Basta NE, Chao DL, Matrajt L, et~al.
\newblock The transmissibility and control of pandemic influenza A (H1N1)
  virus.
\newblock Science. 2009;326(5953):729--733.

\bibitem{cauchemez2009household}
Cauchemez S, Donnelly CA, Reed C, Ghani AC, Fraser C, Kent CK, et~al.
\newblock Household transmission of 2009 pandemic influenza {A (H1N1)} virus in
  the {U}nited {S}tates.
\newblock New England Journal of Medicine. 2009;361(27):2619--2627.

\bibitem{morgan2010household}
Morgan OW, Parks S, Shim T, Blevins PA, Lucas PM, Sanchez R, et~al.
\newblock Household transmission of pandemic {(H1N1)} 2009, {S}an {A}ntonio,
  {T}exas, {USA}, {A}pril-{M}ay 2009.
\newblock Emerg Infect Dis. 2010;16(4):631--637.

\bibitem{france2010household}
France AM, Jackson M, Schrag S, Lynch M, Zimmerman C, Biggerstaff M, et~al.
\newblock Household transmission of 2009 influenza {A (H1N1)} virus after a
  school-based outbreak in {N}ew {Y}ork {C}ity, {A}pril-{M}ay 2009.
\newblock Journal of Infectious Diseases. 2010;201(7):984--992.

\bibitem{carcione2011secondary}
Carcione D, Giele C, Goggin L, Kwan KS, Smith D, Dowse G, et~al.
\newblock Secondary attack rate of pandemic influenza {A (H1N1)} 2009 in
  {W}estern {A}ustralian households, 29 May-7 August 2009.
\newblock Euro Surveill. 2011;16(3):19765.

\bibitem{savage2011assessing}
Savage R, Whelan M, Johnson I, Rea E, LaFreniere M, Rosella LC, et~al.
\newblock Assessing secondary attack rates among household contacts at the
  beginning of the influenza {A (H1N1)} pandemic in {O}ntario, {C}anada,
  {A}pril-{J}une 2009: A prospective, observational study.
\newblock BMC Public Health. 2011;11(1):234.

\bibitem{ng2017association}
Ng S, Saborio S, Kuan G, Gresh L, Sanchez N, Ojeda S, et~al.
\newblock Association between Haemagglutination inhibiting antibodies and
  protection against clade 6B viruses in 2013 and 2015.
\newblock Vaccine. 2017;35(45):6202--6207.

\bibitem{drosten2014transmission}
Drosten C, Meyer B, M{\"u}ller MA, Corman VM, Al-Masri M, Hossain R, et~al.
\newblock Transmission of {MERS}-coronavirus in household contacts.
\newblock New England Journal of Medicine. 2014;371(9):828--835.

\bibitem{arwady2016middle}
Arwady MA, Alraddadi B, Basler C, Azhar EI, Abuelzein E, Sindy AI, et~al.
\newblock Middle East respiratory syndrome coronavirus transmission in extended
  family, Saudi Arabia, 2014.
\newblock Emerging infectious diseases. 2016;22(8):1395.

\bibitem{fang2016transmission}
Fang LQ, Yang Y, Jiang JF, Yao HW, Kargbo D, Li XL, et~al.
\newblock Transmission dynamics of Ebola virus disease and intervention
  effectiveness in Sierra Leone.
\newblock Proceedings of the National Academy of Sciences.
  2016;113(16):4488--4493.

\bibitem{glynn2018variability}
Glynn JR, Bower H, Johnson S, Turay C, Sesay D, Mansaray SH, et~al.
\newblock Variability in intrahousehold transmission of Ebola virus, and
  estimation of the household secondary attack rate.
\newblock The Journal of Infectious Diseases. 2018;217(2):232--237.

\bibitem{reichler2018household}
Reichler MR, Bangura J, Bruden D, Keimbe C, Duffy N, Thomas H, et~al.
\newblock Household transmission of Ebola virus: risks and preventive factors,
  Freetown, Sierra Leone, 2015.
\newblock The Journal of infectious diseases. 2018;218(5):757--767.

\bibitem{marsh2018unwelcome}
Marsh Z, Grytdal S, Beggs J, Leshem E, Gastanaduy P, Rha B, et~al.
\newblock The unwelcome houseguest: secondary household transmission of
  norovirus.
\newblock Epidemiology \& Infection. 2018;146(2):159--167.

\bibitem{tsang2018transmissibility}
Tsang TK, Chen TM, Longini~Jr IM, Halloran ME, Wu Y, Yang Y.
\newblock Transmissibility of norovirus in urban versus rural households in a
  large community outbreak in China.
\newblock Epidemiology. 2018;29(5):675.

\bibitem{hoang2019transmission}
Hoang CQ, Nguyen TTT, Ho NX, Nguyen HD, Nguyen AB, Nguyen THT, et~al.
\newblock Transmission and serotype features of hand foot mouth disease in
  household contacts in Dong Thap, Vietnam.
\newblock BMC Infectious Diseases. 2019;19(1):933.

\bibitem{korpe2019case}
Korpe PS, Gilchrist C, Burkey C, Taniuchi M, Ahmed E, Madan V, et~al.
\newblock Case-control study of cryptosporidium transmission in {B}angladeshi
  households.
\newblock Clinical Infectious Diseases. 2019;68(7):1073--1079.

\bibitem{banerjee2019containing}
Banerjee E, Griffith J, Kenyon C, Christianson B, Strain A, Martin K, et~al.
\newblock Containing a measles outbreak in Minnesota, 2017: methods and
  challenges.
\newblock Perspectives in public health. 2019; p. 1757913919871072.

\bibitem{bi2020epidemiology}
Bi Q, Wu Y, Mei S, Ye C, Zou X, Zhang Z, et~al.
\newblock Epidemiology and transmission of COVID-19 in 391 cases and 1286 of
  their close contacts in Shenzhen, China: a retrospective cohort study.
\newblock The Lancet Infectious Diseases. 2020;.

\bibitem{li2020characteristics}
Li W, Zhang B, Lu J, Liu S, Chang Z, Cao P, et~al.
\newblock The characteristics of household transmission of COVID-19.
\newblock Clinical Infectious Diseases.
  2020;doi:{https://doi.org/10.1093/cid/ciaa450}.

\bibitem{dietz1988first}
Dietz K.
\newblock The first epidemic model: a historical note on PD En'ko.
\newblock Australian Journal of Statistics. 1988;30(1):56--65.

\bibitem{becker1989analysis}
Becker NG.
\newblock Analysis of Infectious Disease Data. vol.~33.
\newblock CRC Press; 1989.

\bibitem{greenwood1931statistical}
Greenwood M.
\newblock On the statistical measure of infectiousness.
\newblock Epidemiology \& Infection. 1931;31(3):336--351.

\bibitem{giesecke2014primary}
Giesecke J.
\newblock Primary and index cases.
\newblock The Lancet. 2014;384(9959):2024.

\bibitem{lau2015inferring}
Lau MS, Cowling BJ, Cook AR, Riley S.
\newblock Inferring influenza dynamics and control in households.
\newblock Proceedings of the National Academy of Sciences.
  2015;112(29):9094--9099.

\bibitem{panum1940observations}
Panum PL, Petersen JJ.
\newblock Observations Made During the Epidemic of Measles on the Faroe Islands
  in the Year 1846.
\newblock Delta Omega Society New York; 1940.

\bibitem{longini2005containing}
Longini IM, Nizam A, Xu S, Ungchusak K, Hanshaoworakul W, Cummings DA, et~al.
\newblock Containing pandemic influenza at the source.
\newblock Science. 2005;309(5737):1083--1087.

\bibitem{kemper1980error}
Kemper JT.
\newblock Error sources in the evaluation of secondary attack rates.
\newblock American Journal of Epidemiology. 1980;112(4):457--464.

\bibitem{RampeyLongini1992}
Rampey AH Jr, Longini IM Jr, Haber M, Monto AS.
\newblock A discrete-time model for the statistical analysis of infectious
  disease incidence data.
\newblock Biometrics. 1992;48:117--128.

\bibitem{kenah2011contact}
Kenah E.
\newblock Contact intervals, survival analysis of epidemic data, and estimation
  of $R_0$.
\newblock Biostatistics. 2011;12(3):548--566.

\bibitem{kenah2013non}
Kenah E.
\newblock Non-parametric survival analysis of infectious disease data.
\newblock Journal of the Royal Statistical Society: Series B (Statistical
  Methodology). 2013;75(2):277--303.

\bibitem{kenah2015semiparametric}
Kenah E.
\newblock Semiparametric relative-risk regression for infectious disease
  transmission data.
\newblock Journal of the American Statistical Association.
  2015;110(509):313--325.

\bibitem{bollobas1998random}
Bollob{\'a}s B.
\newblock Random Graphs.
\newblock Springer; 1998.

\bibitem{RN1}
Gilbert EN.
\newblock Random graphs.
\newblock The Annals of Mathematical Statistics. 1959;30(4):1141--1144.

\bibitem{Python3}
Van~Rossum G, Drake FL.
\newblock Python 3 Reference Manual.
\newblock Scotts Valley, CA: CreateSpace; 2009.

\bibitem{Rlang}
{R Core Team}. R: A Language and Environment for Statistical Computing; 2020.
\newblock Available from: \url{https://www.R-project.org/}.

\bibitem{cameron2011robust}
Cameron AC, Gelbach JB, Miller DL.
\newblock Robust inference with multiway clustering.
\newblock Journal of Business \& Economic Statistics. 2011;29(2):238--249.

\bibitem{liang1986longitudinal}
Liang KY, Zeger SL.
\newblock Longitudinal data analysis using generalized linear models.
\newblock Biometrika. 1986; p. 13--22.

\bibitem{kenah2019handbook}
Kenah E.
\newblock Pairwise survival analysis of infectious disease transmission data.
\newblock In: Held L, Hens N, D~O'Neill P, Wallinga J, editors. Handbook of
  Infectious Disease Data Analysis. CRC Press; 2019. p. 221--244.

\bibitem{sharker2019pairwise}
Sharker Y, Kenah E.
\newblock Pairwise accelerated failure time models for infectious disease
  transmission with external sources of infection.
\newblock arXiv preprint arXiv:190104916. 2019;.

\bibitem{lipsitch2020defining}
Lipsitch M, Swerdlow DL, Finelli L.
\newblock Defining the epidemiology of Covid-19—--studies needed.
\newblock New England Journal of Medicine. 2020;382(13):1194--1196.

\bibitem{o1999bayesian}
O’Neill PD, Roberts GO.
\newblock Bayesian inference for partially observed stochastic epidemics.
\newblock Journal of the Royal Statistical Society: Series A.
  1999;162(1):121--129.

\bibitem{morozova2018risk}
Morozova O, Cohen T, Crawford FW.
\newblock Risk ratios for contagious outcomes.
\newblock Journal of The Royal Society Interface. 2018;15(138):20170696.

\bibitem{Yang-resampling-2007}
Yang Y, Longini J Ira~M, Halloran ME.
\newblock A resampling-based test to detect person-to-person transmission of
  infectious diseases.
\newblock Annals of Applied Statistics. 2007;1:211--228.

\bibitem{Yang-EM-2012}
Yang Y, Longini J Ira~M, Halloran ME, Obenchain V.
\newblock A hybrid {EM} and {M}onte {C}arlo {EM} algorithm and its application
  to analysis of transmission of infectious diseases.
\newblock Biometrics. 2012;68:1238--1249.

\bibitem{rojas2016epidemiology}
Rojas DP, Dean NE, Yang Y, Kenah E, Quintero J, Tomasi S, et~al.
\newblock The epidemiology and transmissibility of Zika virus in Girardot and
  San Andres island, Colombia, September 2015 to January 2016.
\newblock Eurosurveillance. 2016;21(28):30283.

\bibitem{Rgee}
to~R~by Thomas~Lumley VJCP, src/dgedi f BRF, src/dgefa f are for LINPACK
  authored by Cleve Moler Note that maintainers are not available to give
  advice on using a package they did~not author. gee: Generalized Estimation
  Equation Solver; 2019.
\newblock Available from: \url{https://CRAN.R-project.org/package=gee}.

\bibitem{Rreticulate}
Ushey K, Allaire J, Tang Y. reticulate: Interface to 'Python'; 2020.
\newblock Available from: \url{https://CRAN.R-project.org/package=reticulate}.

\bibitem{Rsurvival}
Therneau TM. A Package for Survival Analysis in R; 2020.
\newblock Available from: \url{https://CRAN.R-project.org/package=survival}.

\bibitem{Rtranstat}
Kenah E, Yang Y. transtat: Statistical Methods for Infectious Disease
  Transmission; 2020.
\newblock Available from: \url{https://github.com/ekenah/transtat}.

\bibitem{hagberg2008exploring}
Hagberg A, Swart P, S~Chult D.
\newblock Exploring network structure, dynamics, and function using NetworkX.
\newblock Los Alamos National Lab.(LANL), Los Alamos, NM (United States); 2008.

\bibitem{oliphant2006guide}
Oliphant TE.
\newblock A guide to NumPy. vol.~1.
\newblock Trelgol Publishing USA; 2006.

\bibitem{mckinney-proc-scipy-2010}
{W}es {M}c{K}inney.
\newblock {D}ata {S}tructures for {S}tatistical {C}omputing in {P}ython.
\newblock In: {S}t\'efan van~der {W}alt, {J}arrod {M}illman, editors.
  {P}roceedings of the 9th {P}ython in {S}cience {C}onference; 2010. p. 56 --
  61.

\bibitem{Rmass}
Venables WN, Ripley BD.
\newblock Modern Applied Statistics with S.
\newblock 4th ed. New York: Springer; 2002.
\newblock Available from: \url{http://www.stats.ox.ac.uk/pub/MASS4}.

\bibitem{Rsandwich}
Zeileis A.
\newblock Object-Oriented Computation of Sandwich Estimators.
\newblock Journal of Statistical Software. 2006;16(9):1--16.
\newblock doi:{10.18637/jss.v016.i09}.

\end{thebibliography}

\end{document}